\documentclass[12pt,preprint]{aastex}

\usepackage{natbib}

\newcommand{\AX}{A$^{2}\Delta$--X$^{2}\Pi$}
\newcommand{\CHAX}{CH~A$^{2}\Delta$--X$^{2}\Pi$}
\newcommand{\BX}{B$^{2}\Sigma$--X$^{2}\Sigma$}

\newcommand{\muBohr}{\mu_{\mathrm{Bohr}}}
\newcommand{\gLande}{g^{\mathrm{L}}}

\begin{document}
\title{The Zeeman effect in the G band}

\author{H.\ Uitenbroek}
\affil{National Solar Observatory/Sacramento Peak\footnote{Operated by the %
Association of Universities for Research in Astronomy, Inc. (AURA), %
for the National Science Foundation}, P.O.\ Box 62, Sunspot, NM 88349}
\email{huitenbroek@nso.edu}
\author{E.\ Miller--Ricci}
\affil{Harvard University, 60 Garden Street, Cambridge, MA 02174}
\email{emillerricci@cfa.harvard.edu}
\author{A.\ Asensio Ramos, J.\ Trujillo Bueno\footnote{Consejo Superior
de Investigaciones Cient\'\i ficas, Spain}}
\affil{Instituto de Astrof\'\i sica de Canarias, 38205,
La Laguna, Tenerife, Spain}
\email{aasensio@ll.iac.es,jtb@ll.iac.es}


\begin{abstract}
We investigate the possibility of measuring magnetic field strength
in G-band bright points through the analysis of Zeeman polarization
in molecular CH lines.
To this end we solve the equations of polarized radiative transfer
in the G band through a standard plane-parallel model of the solar
atmosphere with an imposed magnetic field,
and through a more realistic snapshot from a simulation of solar
magneto-convection.
This region of the spectrum is crowded with many atomic and
molecular lines.
Nevertheless, we find several instances of isolated
groups of CH lines that are predicted to produce a measurable
Stokes V signal in the presence of magnetic fields.
In part this is possible because the effective Land\'{e} factors of
lines in the stronger main branch of the \CHAX\ transition tend to zero
rather quickly for increasing total angular momentum $J$, resulting
in a Stokes $V$ spectrum of the G band that is less crowded than the
corresponding Stokes $I$ spectrum.
We indicate that, by contrast, the effective Land\'{e} factors of
the $R$ and $P$ satellite sub-branches of this
transition tend to $\pm 1$ for increasing $J$.
However, these lines are in general considerably weaker, and do
not contribute significantly to the polarization signal.
In one wavelength location near 430.4 nm the overlap of several magnetically
sensitive and non-sensitive CH lines is predicted to result in
a single-lobed Stokes $V$ profile, raising the possibility of
high spatial-resolution narrow-band polarimetric imaging.
In the magneto-convection snapshot we find circular polarization signals
of the order of 1\% prompting us to conclude that measuring magnetic field
strength in small-scale elements through the Zeeman effect in
CH lines is a realistic prospect.
\end{abstract}



\section{Introduction\label{sec:introduction}}
Apart from convective phenomena like the granulation most features we
observe in the solar atmosphere are due to the presence of magnetic fields.
The dynamics of this magnetic field lie at the root of solar activity,
giving rise to variable amounts of UV, X-ray and particle radiation
that influence not only Earth climate, but also our daily
lives in a society that depends more and more on technology.
Thus, there is a real incentive to understand, and eventually
be able to predict, the behavior of the solar magnetic field.
While larger scale magnetic field concentrations like sunspot are more
evident, longer lived, and easier to observe, the field on the solar
surface seems to be distributed predominantly in
small-scale magnetic elements like micropores, network bright points
and low contrast internetwork fields
        \citep{Stenflo1994}.

Because of their small size, characteristically $0''\!\!.25$
	\citep{Muller1985,Berger_etal1995,Berger+Title2001},
at or below the resolution limit of current solar instrumentation,
and their rapid time scale of evolution of 10-100 sec,
near sustainable cadence of multi-wavelength observations,
the morphology, thermal structure and dynamics of small-scale flux
concentrations is difficult to study.
Despite a characteristic field strength in the kilo Gauss range
	\citep{Stenflo1973},
the small sizes of magnetic elements result in low values of the
magnetic flux and make the field difficult to observe.
Therefore, long integration times are needed to reach the required
magnetograph sensitivity if the elements remain unresolved,
and this results in smeared images due to the short dynamical
time scales of the elements.
Fortunately, indirect methods which allow much shorter
exposure times exist to examine the structure of magnetic elements,
and to follow their evolution.
In particular, these methods involve imaging the Sun in wideband
(of order 1 nm) filters centered on molecular band heads towards
the blue end of the spectrum.
Examples are the band head at 388.3 nm due to \BX\ electronic
transitions of CN
	\citep{Sheeley1971},
and the so-called G band
	\citep[originally designated by][]{Fraunhofer1817}
around 430.5 nm due to \AX\ transitions of the CH molecule
	\citep{Muller+Hulot+Roudier1989,Muller+Roudier1992,%
Muller+Roudier+Vigneau+Auffret1994,Berger_etal1995,%
VanBallegooijen_etal1998,Berger+Title2001}.

In these molecular wideband images the magnetic elements appear as
subarcsecond sized bright points with contrasts of typically 30\%,
compared to the average photosphere
	\citep{Berger_etal1995}.
	\citet{Berger+Title2001}
find that G-band bright points are cospatial and comorphous
with magnetic elements in intergranular lanes to within $0.''\!\!24$.
Also in strong photospheric lines
	\citep{Chapman+Sheeley1968,Spruit+Zwaan1981}
and the wings of chromospheric lines like H$\alpha$
	\citep{Dunn+Zirker1973},
and \ion{Ca}{2} K
	\citep{Mehltretter1974}
bright points seem to coincide with magnetic field concentrations.
However, the cores of these lines are considerably darker than
the integrated G-band intensity and require much narrower bandpasses,
making imaging in them much less practical.
Until now measurement of magnetic field strength in molecular
bright points has had to rely on cospatial magnetograms in
magnetically sensitive atomic lines.
This is an arduous process, which involves careful alignment
of images taken at different wavelengths and a relative ``destretching''
of images taken at different times to prevent contamination
of the magnetogram signal due to seeing induced image distortions.
In addition, not every magnetic element has an associated bright
point or vice versa, at a given time, although both
	\citet{Muller_etal2000} and
	\citet{Berger+Title2001}
conclude that there is a one-to-one correspondence when the temporal
evolution of the magnetic field and bright grains is followed.

An important point in the diagnostic use of bright point
imaging for the study of magnetic element dynamics and morphology is
the question of why the associated bright points are
particularly bright in the molecular band heads.
Model calculations of G-band brightness in semi-empirical fluxtube
atmospheres provide reasonable values for the bright point contrast
\citep{SanchezAlmeida+AsensioRamos+TrujilloBueno+Cernicharo2001,
Rutten+Kiselman+Rouppe+Plez2001,Steiner+Hauschildt+Bruls2001}.
Flux-tube models are hotter than the quiet Sun in the
layers producing the observed light. They are evacuated owing to the
presence of a magnetic field, thus allowing us to observe deep and
therefore hot photospheric layers. For this very reason,
\citet{SanchezAlmeida+AsensioRamos+TrujilloBueno+Cernicharo2001}
concluded that G-band contrast in the bright points is enhanced compared
to the surrounding photosphere because the opacity in the CH lines
is less affected by the higher temperatures present in magnetic elements
than the continuum opacity, which is mostly due to H$^-$.
In the continuum a rise in temperature leads to a higher
formation height of intensity at consequently lower temperatures.
On the other hand, in the G band the formation height remains more
or less constant (because of the increase in H$^-$ opacity and decrease
in CH line opacity through dissociation) so that the emergent intensity
reflects the higher temperatures. Similarly,
\citet{Rutten+Kiselman+Rouppe+Plez2001}
emphasized that the evacuation of magnetic elements due to the requirement
of pressure balance with the non-magnetic surroundings leads to excess
CH dissociation making the G band relatively more transparent,
thereby exposing hotter layers below. This has been clearly illustrated by
\citet{Uitenbroek2003} by modeling the G-band intensity in a
snapshot from a three-dimensional simulation of solar magneto-convection.
That the CH concentration
is strongly reduced in evacuated flux-tube regions is logical because
the molecular concentration depends quadratically on density {\em if} the
molecular dissociation is not controlled by radiation,
which seems to be indeed the case as pointed out by
\citet{SanchezAlmeida+AsensioRamos+TrujilloBueno+Cernicharo2001}.
Such a quadratic dependence on density would
explain why CH lines show stronger
weakening than atomic lines in the G band
    \citep{Langhans+Schmidt+Rimmele+Sigwarth2001},
and cause higher bright-point contrast in the CH and CN band heads
than in atomic line cores, because in the solar photosphere atomic
ionization is mostly dominated by the radiation field,
making the opacity in atomic lines only linearly dependent on the
density in the flux concentrations.

In this paper we propose a technique that hopefully will allow
us to shed further light on the relationship between the magnetic field
in the bright points and their increased contrast,
namely the measurement of the field through analysis of the polarization
signal due to the Zeeman effect in the CH lines that make up the bulk
of the opacity in the G band.
Over the last few years, we have witnessed an increasing interest
in using molecular line polarization as a tool for magnetic field
diagnostics, concerning both the molecular Zeeman effect
        \citep[e.g.,][]{Berdyugina+Frutiger+Solanki+Livingston2000,%
Berdyugina+Solanki2002,AsensioRamos+TrujilloBueno2003}
and the Hanle effect in molecular lines
        \citep[e.g.,][]{LandiDeglInnocenti2003,TrujilloBueno2003b}.
The quantum theoretical background for treating the molecular Zeeman
effect has been mostly in place since 1930
	\citep[see][for a description and references]{Herzberg1950},
but was first worked out practically in terms of splitting patterns
and intensity of polarization components for Hund's case (a) transitions
of any multiplicity and Hund's case (b) transitions of doublets by	
	\citet{Schadee1978}.
His results were extended by
	\citet{Berdyugina+Solanki2002}
to general multiplicity in Hund's case (b) and transitions
of intermediate (a-b) coupling, 
i.e., with partial spin decoupling.
Here we assume the upper and lower states of the \AX\ transitions of the
CH molecule to be in pure Hund's case (b), where the electron spin is
decoupled from the molecular rotation
        \citep[e.g.,][p.\ 303]{Herzberg1950}.
This is a good approximation for these states because the of small
values of the spin coupling constants in both,
as discussed by
        \citet{Berdyugina+Solanki2002}.
These authors explicitly show the difference in effective Land\'{e} factor
between case (b) and the intermediate case (a-b, their Fig.\ 9).
Even for very small rotational quantum numbers the perturbations caused
by incomplete spin decoupling are very small, justifying our assumption.

The structure of this paper is as follows.
In Section \ref{sec:Zeeman} we briefly review the Zeeman effect
in CH lines and show the effective Land\'{e} factors of satellite
branch transitions, which behave differently than those in the main branch.
We discuss our model calculations in Section \ref{sec:modelcalculations},
and give conclusions in Section \ref{sec:conclusions}.

\section{The Zeeman effect in lines of the CH molecule\label{sec:Zeeman}}
Radiative transitions in a molecule are susceptible to the Zeeman effect in
the presence of an external magnetic field and may produce polarized
radiation that can be analyzed to infer the properties of the field,
much like in the atomic case.
If a molecule has a non-zero magnetic moment, the interaction of that
moment with the magnetic field will split the molecular energy levels
into a pattern that is in general, however, more complex than in the
atomic case because of the additional degrees of freedom inherent in
molecular structure.
In a diatomic molecule both vibration along the internuclear axis and
rotation around the center of mass influence level energies, but
only rotation affects the molecule's magnetic moment because it
alters the way the electronic spin and orbital angular momenta 
add to the total angular momentum $\mathbf{J}$ of the molecule,
which in turn couples to the magnetic field.

\subsection{Energy level splitting}
Three factors can contribute to the magnetic moment of a diatomic molecule:
the orbital and spin angular momenta of the electrons, the rotation of
the molecule, and the magnetic moment associated with the nuclear spins.
The first contribution is of the order of a Bohr magneton
$\muBohr = (e B/2 m_{\mathrm{e}}) \hbar$ (in SI units), while the
second and third contributions are smaller by a factor of the order 
$m_{\mathrm{p}} / m_{\mathrm{e}} \approx 1850$.
Since the CH lines we consider in this paper have non-zero orbital
and spin angular momenta in both their upper and lower levels,
we can safely ignore the contributions of molecular rotation and
nuclear spin here.

Let $\mathbf{L}$ and $\mathbf{S}$ be the total electronic orbital and
spin angular momenta in the diatomic molecule, and $\mathbf{\Lambda}$
the component of $\mathbf{L}$ along the internuclear axis.
If $\mathbf{\Lambda}$ is non-zero it is coupled to the electric
field along the internuclear axis, and is space quantized with
quantum number $\Lambda = 0, 1, 2, 3, \ldots$.
In spectroscopic notation these states are denoted
by the upper-case Greek letters $\Sigma$, $\Pi$, $\Delta$, $\Phi$,
etc., which are similar to the $S$, $P$, $D$, $F$, designations in
atomic systems.
The total angular momentum $\mathbf{J}$ is always the resultant
of $\mathbf{L}$, $\mathbf{S}$, the angular momentum of
rotation $\mathbf{R}$, and the total spin of the nuclei.
Different simplifying cases can be distinguished depending on how
strong the relative coupling between these angular momenta is,
as first put forward by Hund
  \citep[see][p.\ 218]{Herzberg1950}.
If both spin and orbital angular momenta are strongly coupled to the
internuclear axis and only very weakly to the rotation,
Hund's case (a) applies.
If on the other hand $\mathbf{S}$ is only coupled weakly to
the internuclear axis, whereas $\mathbf{\Lambda}$ is strongly
coupled we have Hund's case (b).
This is the case that applies to both the upper and lower level of the
CH lines in the G-band region and is discussed further here.
\begin{figure}[htb]
  \epsscale{0.50}
  \plotone{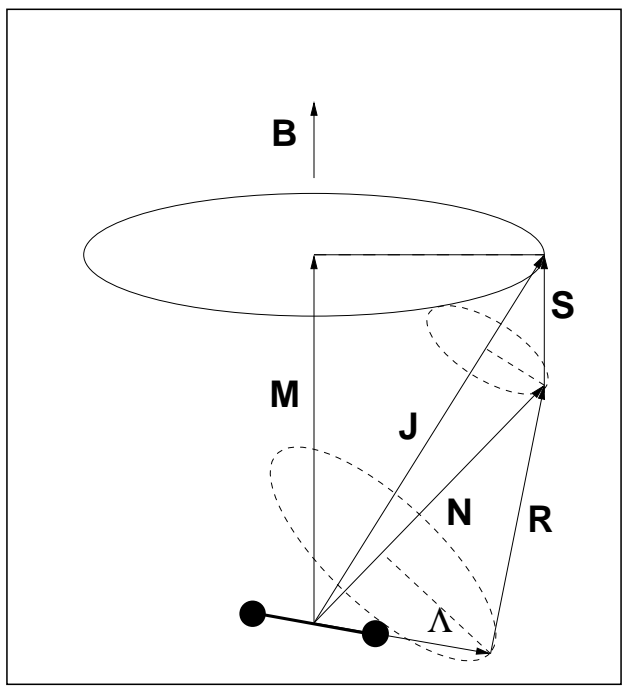}
  \caption[]{Vector diagram of angular momenta in a diatomic
             molecule in the presence of a magnetic field $\mathbf{B}$
             for Hund's case (b) of weak
             coupling of the spin angular momentum $\mathbf{S}$ to
             the internuclear axis.\label{fig:Hunds_b}}
  \epsscale{1.0}
\end{figure}
Figure \ref{fig:Hunds_b} shows a vector diagram of the angular momenta
in Hund's case (b).
In this case the orbital angular momentum $\mathbf{\Lambda}$ and rotational
angular momentum $\mathbf{R}$ combine to form $\mathbf{N}$,
with quantum number
\begin{equation}
  N = \Lambda + R; \qquad R = 0, 1, 2, 3, \ldots.
\end{equation}
$\mathbf{N}$ in turn combines with spin angular momentum $\mathbf{S}$
to form $\mathbf{J}$, which can have quantum numbers
\begin{equation}
  J = N+S, N+S-1, \ldots, |N-S|,
\end{equation}
according to the standard addition rule for quantized angular momentum vectors
  \citep[e.g.,][p.\ 234]{Slater1960}.

In a magnetic field the total angular momentum $\mathbf{J}$ is space
quantized such that the component in the field direction is $M\hbar$,
where
\begin{equation}
  M = J, J-1, \ldots, -J.
\end{equation}
In case the molecule has a non-zero magnetic moment a precession of
$\mathbf{J}$ around the field will take place, and states with different
$M$ will have different energies:
\begin{equation}
  E = E_0 - \overline{\mu}_{B} B =
    E_0 - \overline{\mu}_{J} \cos(\mathbf{J}, \mathbf{B}) B,
  \label{eq:energydiff}
\end{equation}
where $\overline{\mu}_{B}$ is the mean value of the component of
the molecular magnetic moment along the field direction,
$\overline{\mu}_{J}$ is the mean magnetic moment along the direction of
$\mathbf{J}$, and $B$ is the field strength.
The time average of $\overline{\mu}_{J}$ is composed of a contribution
$\Lambda \muBohr$ due to the orbital angular momentum
which is coupled to the internuclear axis in Hund's case (b),
and a contribution $2 \sqrt{S(S+1)} \muBohr$ due to the
spin coupled to the rotation axis.
Since we are considering field strengths outside the Paschen-Back
regime the precession of $\mathbf{J}$ around the field can be considered
slow compared to that of $\mathbf{N}$ and $\mathbf{S}$ about $\mathbf{J}$,
while in Hund's case (b) the latter is in turn slower than the
nutation of $\mathbf{\Lambda}$ about $\mathbf{N}$.
Therefore, we have
\begin{eqnarray}
  \overline{\mu}_{B} &=& \left[ \Lambda \cos(\mathbf{\Lambda}, \mathbf{N})
                         \cos(\mathbf{N}, \mathbf{J}) +
                         2 \sqrt{S(S+1)} \cos(\mathbf{S}, \mathbf{J}) \right]
                         \cos(\mathbf{J}, \mathbf{B}) \muBohr\\
                     &=& \left[ \frac{\Lambda^2}{\sqrt{N(N+1)}}
                         \cos(\mathbf{N}, \mathbf{J}) +
			 2 \sqrt{S(S+1)} \cos(\mathbf{S}, \mathbf{J}) \right]
                         \frac{M}{\sqrt{J(J+1)}} \muBohr,
  \label{eq:meanmuB}
\end{eqnarray}
where we used $\cos(\mathbf{\Lambda}, \mathbf{N}) =
\Lambda / \sqrt{N(N+1)}$ and 
$\cos(\mathbf{J}, \mathbf{B}) = M / \sqrt{J(J+1)}$
  \citep[][p.\ 303]{Herzberg1950}.

According to equation (\ref{eq:meanmuB}) the magnetic field splits
the molecular level (for given $\mathbf{N}$ and $\mathbf{S}$) into
$2J + 1$ equidistant components with a splitting that only depends
on the relevant quantum numbers, not on the molecular constants.
Closer inspection also reveals that when $N$ is large and the first
term in the bracket vanishes, the result becomes independent of $J$
for large $J$ and the maximum splitting (i.e., $M = \pm J$) is of the
order of the normal Zeeman splitting.
This is the case, for instance, when the rotational quantum number $R$
is large so that $\mathbf{\Lambda}$ is almost perpendicular to
$\mathbf{J}$ (see Figure \ref{fig:Hunds_b}), resulting in a contribution
to the magnetic moment due to the orbital angular momentum that
averages to nearly zero.
For a given $\mathbf{J}$ and given orientation of $\mathbf{S}$ with
respect to $\mathbf{J}$, each spin level is split into $2N+1$ components
with a splitting that reduces approximately linearly with $N$ for large
values of $N$ (according to the first term in eq.\ [\ref{eq:meanmuB}];
see also fig.\ 145 in \citeauthor{Herzberg1950} \citeyear{Herzberg1950}). 

Using vector rules
\begin{eqnarray}
  \mathbf{J} &=& \mathbf{N} + \mathbf{S} \\ \nonumber
  \mathbf{N} \cdot \mathbf{J}  &=& 
     |\mathbf{N}| |\mathbf{J}| \cos(\mathbf{N}, \mathbf{J}),
\end{eqnarray}
the quantum rule $|\mathbf{N}| = \sqrt{N(N+1)}$, and similar
expressions for $|\mathbf{S}|$ and $|\mathbf{J}|$ it follows that
\begin{equation}
  \cos(\mathbf{N}, \mathbf{J}) = \frac{J(J+1) + N(N+1) - S(S+1)}
    {2\sqrt{J(J+1)}\sqrt{N(N+1)}},
  \label{eq:cosNJ}
\end{equation}
and similarly
\begin{equation}
  \cos(\mathbf{S}, \mathbf{J}) = \frac{J(J+1) - N(N+1) + S(S+1)}
    {2\sqrt{J(J+1)}\sqrt{S(S+1)}}
  \label{eq:cosSJ}
\end{equation}
\citep[][but note the minor error in the denominator of
the latter expression there]{Berdyugina+Solanki2002}.
Combining equations (\ref{eq:energydiff}, \ref{eq:meanmuB}, \ref{eq:cosNJ},
and \ref{eq:cosSJ}) we obtain
\begin{equation}
  \Delta E = E - E_0 = \gLande M \muBohr B,
\end{equation}
with $\gLande$ the Land\'{e} factor for energy-level splitting in Hund's
case (b) given by
\begin{eqnarray}
  \gLande &=& \Biggl\{
    \frac{\Lambda^2 \left[J(J+1) + N(N+1) -
                          S(S+1)\right]}{2N(N+1)} + \\ \nonumber
    && J(J+1) - N(N+1) + S(S+1)\Biggr\} \frac{1}{J(J+1)}.
  \label{eq:Hundsbsplitting}
\end{eqnarray}

\subsection{Line splitting and polarization}
Electronic transitions in the molecule obey selection rules in the same way
as atomic transitions.  Rules for molecular transitions can be split
up into two categories: general selection rules that hold true 
for all electronic transitions and selection rules that only hold true for
specific coupling cases.  

For all electric dipole transitions in molecules the selection rule for
total angular momentum quantum number $J$ holds rigorously:
\begin{equation}
  \Delta J = 0,\pm 1;
    \quad\mathrm{but}\quad J = 0 \not\leftrightarrow J = 0.
\end{equation}
If both the upper and lower levels of a transition are of Hund's
case (a) or (b), then the quantum numbers for the spin angular momentum
quantum number $S$ and orbital angular momentum $\Lambda$ are defined for
both levels and obey the selection rules:
\begin{eqnarray}
  \Delta S       &=& 0 \\ \nonumber
  \Delta \Lambda &=& 0,\pm 1.
\end{eqnarray}
When both upper and lower level are strictly of case (b),
i.e., with the spin coupled to the molecular rotation,
the quantum number $N$ is defined and obeys the selection rule
\begin{equation}
  \Delta N  = 0, \pm 1,
  \label{eq:selectN}
\end{equation}
with the added restriction that $\Delta N = 0$ is not allowed for
$\Sigma$--$\Sigma$ transitions
  \citep[][p.\ 244]{Herzberg1950}.
The set of transitions that have $\Delta N = \Delta J$ are called
main branch transitions.
\begin{figure}[htb]
  \plotone{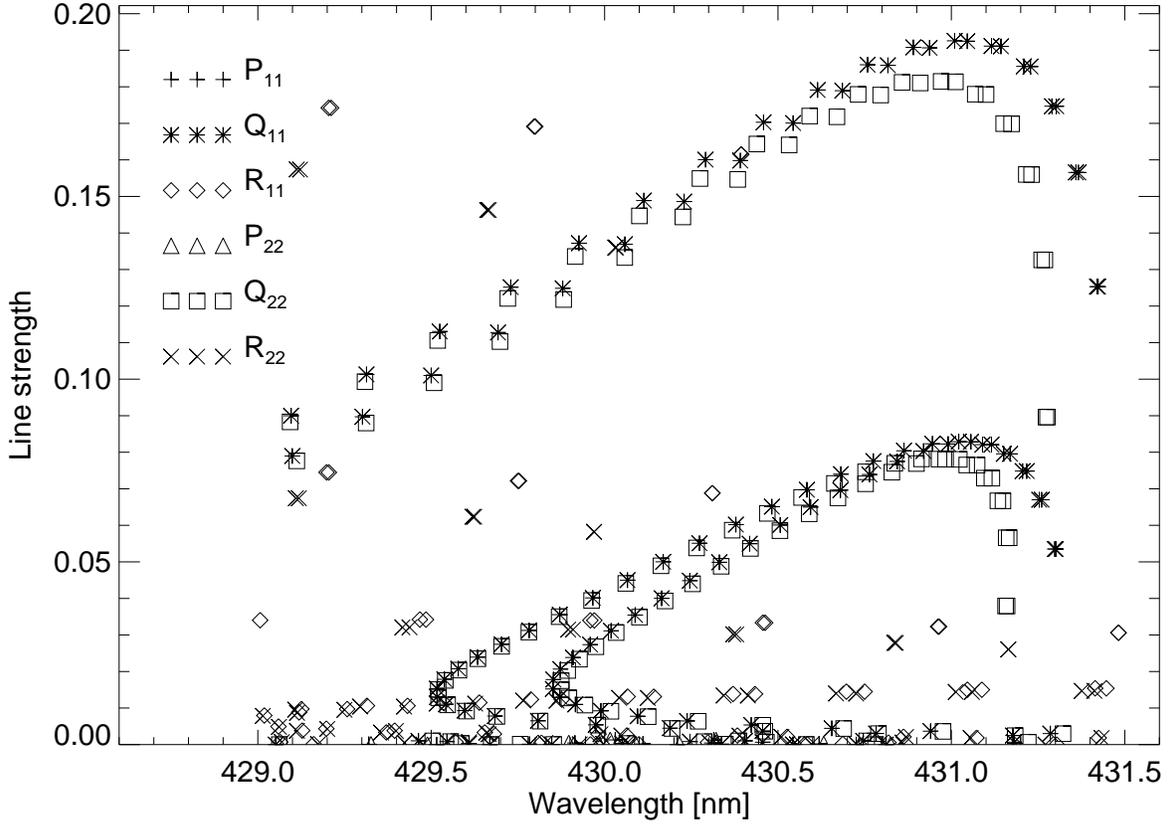}
  \caption[]{Line strengths of the main branches of the \CHAX\
             system in the G-band region for a temperature of $5\times 10^3$~K.
             Values for the oscillator strengths and level energies 
	     were taken from the line list of
	     \citet{Jorgensen+Larsson+Twamae+Yu1996}.
	     In this wavelength region the list includes transitions from
             the first four vibrational bands.
             \label{fig:strength_main}}
\end{figure}
These are termed $R_{ii}$, $Q_{ii}$, and $P_{ii}$ branches for
$\Delta J = J_l - J_u = -1, 0, +1$, respectively, with $J_u$ referring
to the upper level total angular momentum quantum number and $J_l$
to the lower level,
and $i = 1, 2, \ldots$ depending on whether $J = N+S, N+S-1, \ldots$
down to $J = |N-S|$
  \citep[][pp.\ 169 and 222]{Herzberg1950}.
Line strengths (defined as $g_i f \exp(-E_i/kT)$, where
$g_i$ and $E_i$ are the statistical weight and energy of the lower level,
$f$ is the oscillator strength, $k$ the Boltzmann constant, and $T$
the temperature) of \CHAX\ main branch lines in the G-band region
are plotted in Figure~\ref{fig:strength_main} as a function of wavelength
for a typical photospheric temperature of $5\times 10^3$~K.

For $\Delta J \neq \Delta N$, the lines form what are known as the satellite
branches. 
In case (b), the satellite branches lie near the main branches with 
the same values of $\Delta N$, but the line strengths in the satellite
branches are always considerably smaller than those in the main branches
and fall rapidly with increasing $N$
  \citep[][p.\ 244]{Herzberg1950}.
Satellite branches are termed $^{Q}R_{ij}$, $^{P}Q_{ij}$, etc.\ with
the left superscript referring to the value of $\Delta N = N_l - N_u$
in the same way as the main branches are denoted.
\begin{figure}[htb]
  \plotone{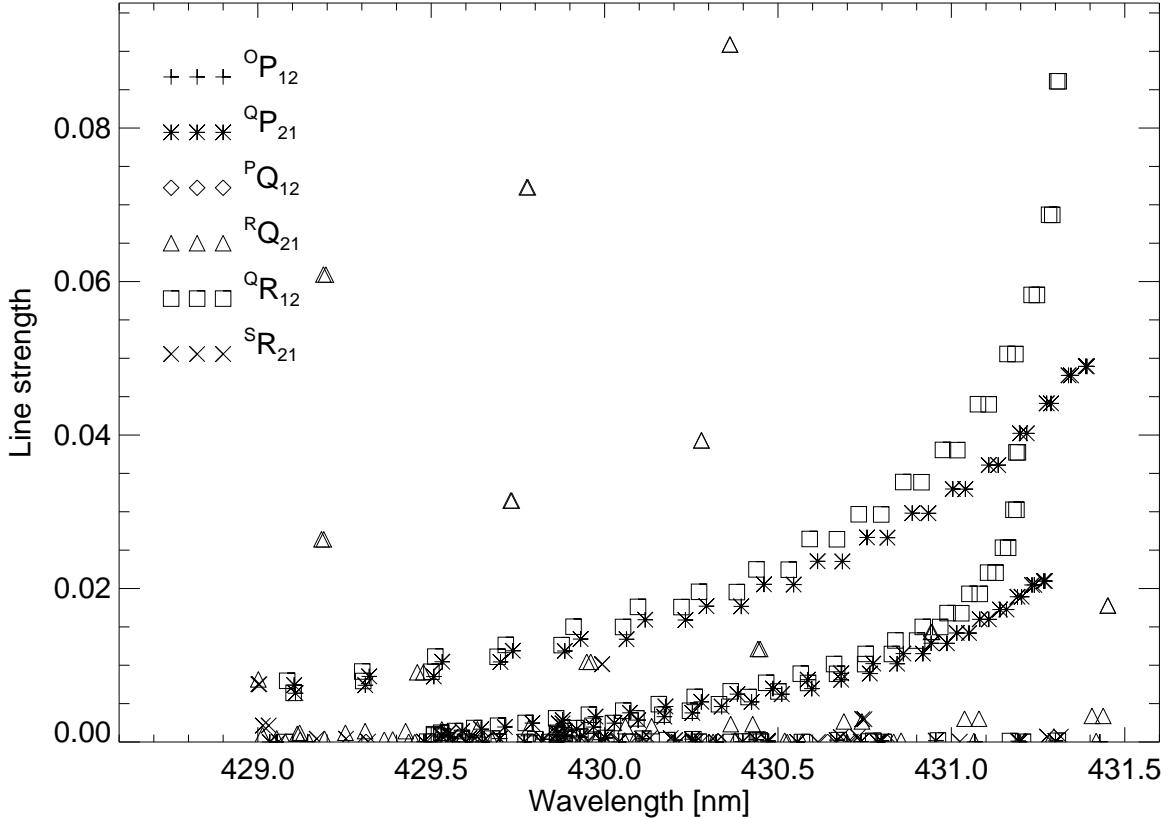}
  \caption[]{Line strengths of the satellite branches of the \CHAX\
             system in the G-band region. \label{fig:strength_sat}}
\end{figure}
Strengths of the \CHAX\ satellite lines in the G-band region are
plotted in Figure~\ref{fig:strength_sat}.
Comparison with Figure~\ref{fig:strength_main} shows that the
satellite branch lines are generally much weaker than those in the
main branch.
Note that Figure~\ref{fig:strength_sat} also contains values for
$^{S}P_{21}$ and $^{O}P_{12}$ branches, in which the superscripts
$S$ and $O$ imply $\Delta N = -2, +2$, respectively.
These values violate selection rule (\ref{eq:selectN}), and indicate that
the associated ``forbidden'' transitions have small but non-zero
transition probabilities.

In the presence of a magnetic field radiative transitions
between shifted levels are allowed with the following selection rule for
the component $M$ of the total angular momentum along the magnetic
field:
\begin{equation}
  \Delta M = 0, \pm 1; \qquad \mathrm{but}\quad
  M = 0 \not\rightarrow M = 0 \quad\mathrm{if}\quad \Delta J = 0.
\end{equation}
These transitions have characteristic polarizations that follow the same
rules as for the atomic case.
In the case of $\Delta M = 0$ the observed line component is polarized 
parallel to the field when viewed perpendicular to the field
($\pi$-component), and in the case of $\Delta M = \pm 1$,
the components are linearly polarized perpendicular to the field
when viewed perpendicular to the field and left- and right hand
circularly polarized when viewed in the field direction
($\sigma_{\pm}$-components).
The shift in energy of each individual line component with respect to
its zero-field value is:
\begin{equation}
  \Delta E = -(\gLande_l M_l - \gLande_u M_u) \muBohr B.
\end{equation}
As for atomic transitions the strength for the individual
components with $M_l \rightarrow M_u$ within each radiative
transition from a lower level with
quantum numbers $(\Lambda_l, S_l, N_l, J_l)$ to an upper level
with $(\Lambda_u, S_u, N_u, J_u)$ is obtained from the
Wigner-Eckart theorem
         \citep{Brink+Satchler1968}.
Table 1 lists the unnormalized strengths $S_q(M_l, M_u)$ for each
allowed $q = M_l - M_u$ and $\Delta J$.
These strengths have to be normalized for each given $q$ according to:
\begin{equation}
  \sum_{M_l,M_u} S_q(M_l, M_u) = 1,
\end{equation}
so that the sum of strengths over all Zeeman split components with a
particular polarization within one transition yields unity.
\begin{table}[tbh]
  \centering
  \textsc{TABLE 1\\
          The unnormalized strengths $S_q(M_l, M_u)$}\\[1.5ex]
  \begin{tabular}{l|ccc}
    \hline\hline
      $J_{l} - J_{u}$  &  $q = 0$   &  $q = -1$  &  $q = +1$ \\ \hline
      $0$   & $2M_{u}^{2}$          &  $(J_u + M_u)(J_u - M_u + 1)$ &
        $(J_u - M_u)(J_u + M_u + 1)$ \\
      $-1$  & $2(J_u^2 - M_u^2)$    &  $(J_u + M_u)(J_u + M_u - 1)$ &
        $(J_u - M_u)(J_u - M_u - 1)$ \\
      $+1$  & $2(J_u + 1)^2 -2M_u^2$ & $(J_u - M_u + 1)(J_u - M_u + 2)$ &
        $(J_u + M_u + 1)(J_u + M_u + 2)$ \\ \hline
  \end{tabular}
  \label{tab:componentstrength}
\end{table}

A convenient quantity to describe the characteristics of an anomalous Zeeman
multiplet is the effective Land\'{e} $g$ factor for the transition,
which corresponds to the $g$-factor of a Zeeman triplet whose
$\sigma$-components lie at the wavelengths of the centers of gravity
of the $\sigma$-components of the anomalously split line.
Thus the effective Land\'{e} factor for a general Zeeman multiplet
is calculated as the normalized and weighted sum of the shift
$\gLande_l M_l - \gLande_u M_u$ over all transitions in the multiplet
with $q = 1$:
\begin{equation}
  g_{\mathrm{eff}} = \frac{
    \sum_{M_u, M_l} S_q(M_l, M_u) (\gLande_l M_l - \gLande_u M_u)}{
    \sum_{M_u, M_l} S_q(M_l, M_u)}.
  \label{eq:effectiveLande}
\end{equation}

\begin{figure}[htb]
  \plotone{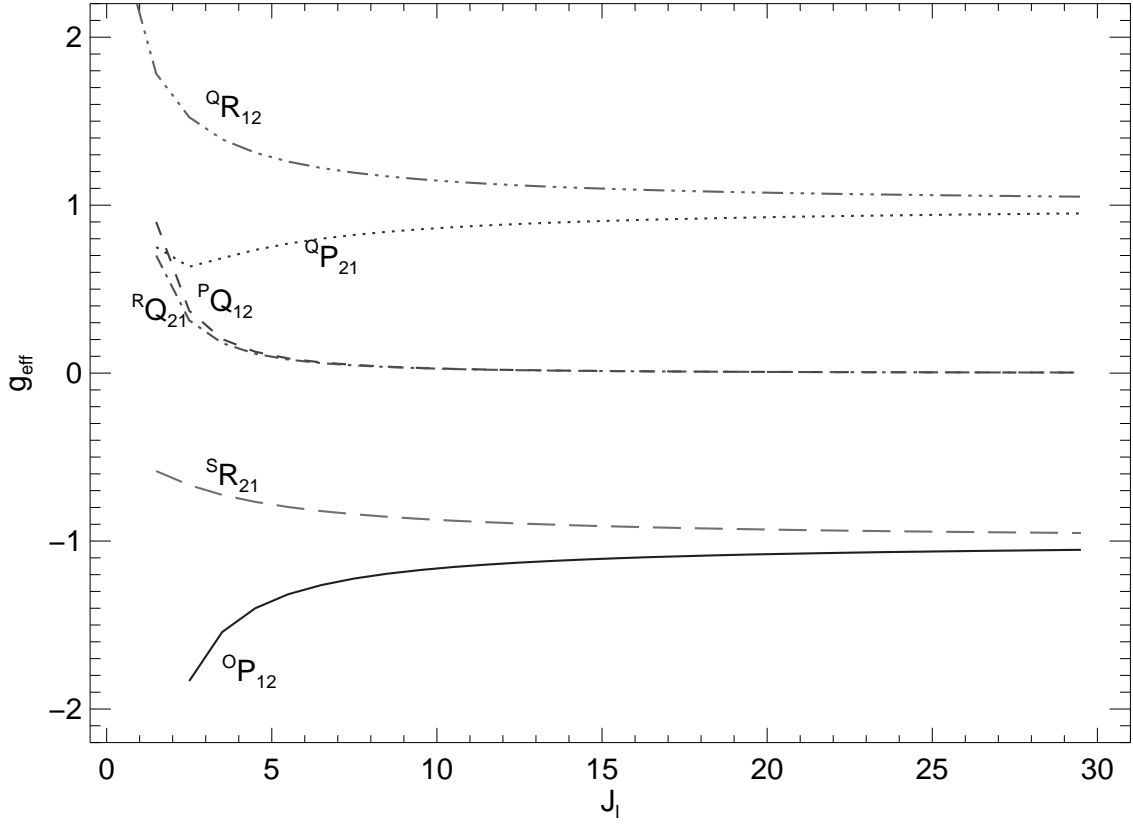}
  \caption[]{Effective Land\'{e} factors of the satellite branches
             of the \CHAX\ system as function of the lower-level
             angular momentum quantum number $J_l$.\label{fig:satellite}}
\end{figure}
Equation \ref{eq:effectiveLande} was used to calculate the effective
Land\'{e} factors of the satellite branches of the \CHAX\ system in
the G-band region as a function of the lower-level angular momentum quantum
number $J_l$ shown in Figure \ref{fig:satellite}.
A similar graph is shown by
   \citet[][their Fig.\ 9]{Berdyugina+Solanki2002}
for the main branch transitions of the CH molecule.
Note that while $g_{\mathrm{eff}}$ for all the main
branches tends to zero with increasing $J_l$, this is not the case
for the those of the $^{Q}P_{21}$ and $^{Q}R_{12}$,
and $^{O}P_{12}$ and $^{S}R_{21}$
type satellite branches, which tend to $+1$ and $-1$, respectively.
Only effective Land\'{e} factors of the $Q$ type satellite branches
follow the behavior of the main branch transitions.
Despite the tendency of $g_{\mathrm{eff}}$ of the $P$ and $R$ satellite
branches to $\pm 1$, we found in the calculations described below
that the satellite branch lines contribute little to the
polarization signal in the G band because they are generally much
weaker than the main branch lines as indicated by comparison of Figures
\ref{fig:strength_main} and \ref{fig:strength_sat}.

\section{Model calculations\label{sec:modelcalculations}}
To get a first impression of the feasibility of bright point magnetic field
measurement through the Zeeman effect in the CH lines of the G band we
calculated the polarization at these wavelengths due to a 0.1 T constant
vertical field in a one-dimensional plane-parallel model of the quiet
solar atmosphere.
In addition, we modeled the polarization signal through a two-dimensional
cross section of a snapshot from a three-dimensional solar magneto-convection
simulation.

\subsection{Radiative transfer}
We implemented the rules for energy level splitting in Hund's case
(b) (see eq.\ [\ref{eq:Hundsbsplitting}]) and line component strength
(Table 1) in a numerical radiative transfer code to calculate the splitting
patterns of the CH lines in the G-band region.
The code accounts for all components in each of the CH lines and
does not rely on the calculated effective Land\'e factors.

Concentration of the CH molecule in each location of the models
was calculated by solving the coupled chemical equilibrium equations
of a set of 12 molecules consisting of H$_2$, H$_2^+$, C$_2$,
N$_2$, O$_2$, CH, CO, CN, NH, NO, OH, and H$_2$O, and their
atomic constituents.
The population of individual CH levels was calculated according
to the Saha--Boltzmann equations, so that the molecular line opacity
and source function were assumed to be in Local Thermodynamic Equilibrium
(LTE).
Values of wavelengths and line strengths were taken from file
HYDRIDES.ASC of CD-ROM 18 by
Kurucz\footnote{\textsf{http://kurucz.cfa.harvard.edu}},
which results in a similar emergent spectrum as the list provided by
        \citet{Jorgensen+Larsson+Twamae+Yu1996},
although the latter includes many more weaker lines.
In addition, opacities due to background atomic lines were included
from files gf0430.10 and gf0440.10 of CD-ROM 1 by Kurucz$^3$.
The opacities and source functions of these lines were assumed to
be in LTE as well, and for each of the electric dipole transitions
among them Zeeman splitting patterns
        \citep[e.g.,][pp.\ 107-111]{Stenflo1994}
were calculated as input for the polarized radiative transfer calculation.

For each wavelength for which the emergent intensity was to be calculated
the contributions from molecular and atomic lines to the $4 \times 4$
propagation matrix $\mathbf{K}$ and the 4-element Stokes emissivity
$\mathbf{e}$
        \citep{TrujilloBueno2003}
were added as well as the (unpolarized) contributions from relevant
background continuum processes. 
Given the propagation matrix and emission vector the Stokes transfer
equation was solved using the quasi-parabolic DELO method proposed by
	\citet[][see \citeauthor{SocasNavarro+TrujilloBueno+RuizCobo2000}
2000 for a first application of this method]{TrujilloBueno2003}.
The DELO method works analogous to the short characteristics method
        \citep{Kunasz+Auer1988,Auer+Fabiani-Bendicho+Trujillo-Bueno1994}
for unpolarized radiation and is easily generalized to multi-dimensional
geometry.

\subsection{Plane-parallel atmosphere with imposed magnetic field}
\begin{figure}[p]
  \epsscale{0.785}
  \plotone{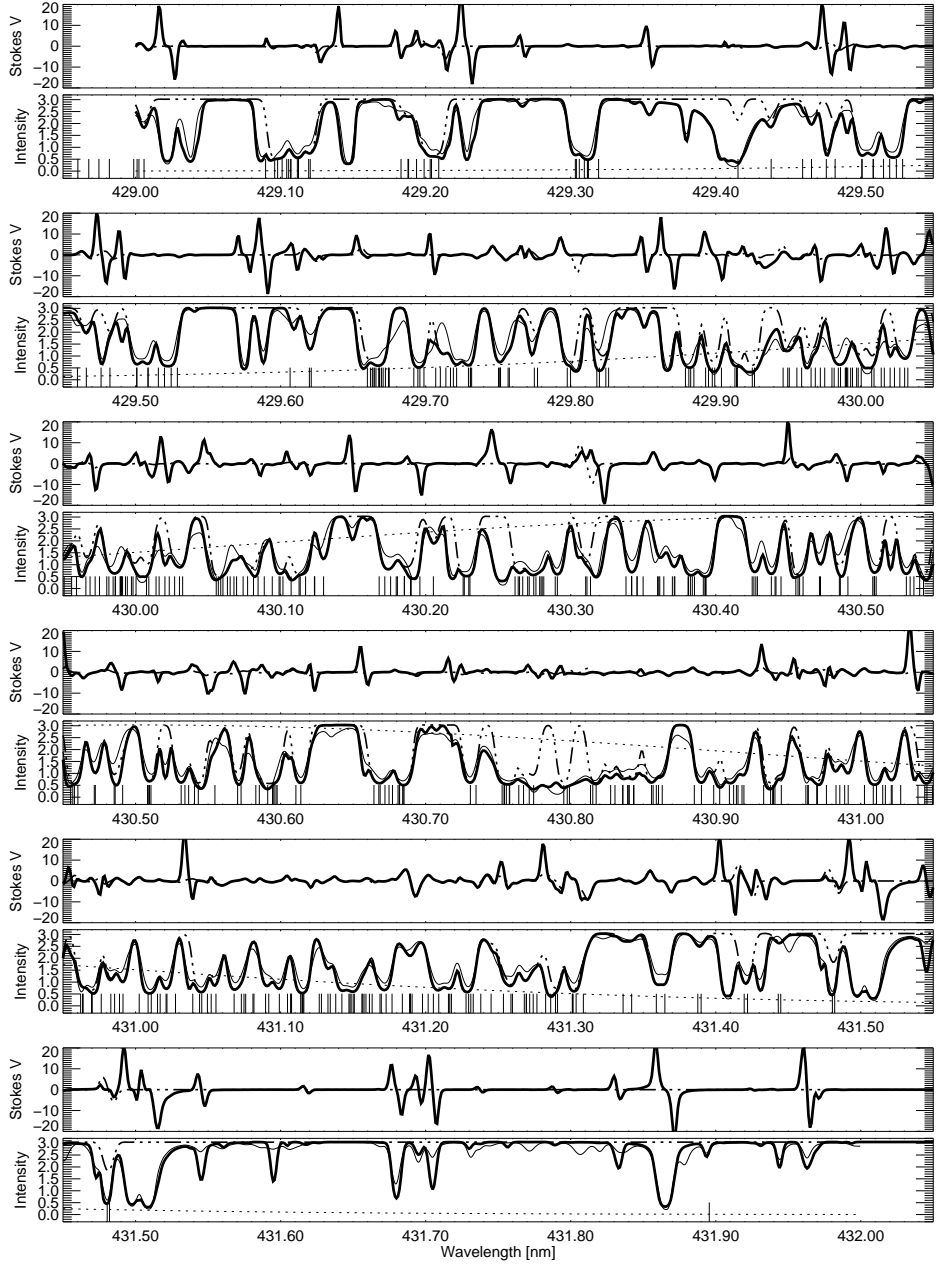}
  \caption[]{Intensity (in units of
    $10^{-8}$ J m$^{-2}$ s$^{-1}$ Hz$^{-1}$ sr$^{-1}$) and Stokes $V$ profiles
    (in percentage of continuum intensity) in the G-band region.
    Thick curve is the calculated spectrum accounting for both atomic and
    molecular CH lines. Dash-dotted line represents spectrum due to CH lines
    only. Thin curve is the central intensity atlas, the dotted
    curve represents a typical G-band filter with 1 nm FWHM,
    and each of the vertical marks indicates the position of a CH line.
    \label{fig:Gbandspectrum}}
  \epsscale{1.0}
\end{figure}
To investigate whether the intensity and polarization spectra
anywhere in the G band are dominated solely by CH lines we solved the
full Stokes radiative transfer equations over a wavelength range
of 3 nm centered around 430.50 nm through a one-dimensional hydrostatic
model of the average quiet solar atmosphere
	\citep[FALC, model C of][]{Fontenla+Avrett+Loeser1993}
with an imposed vertical magnetic field of 0.1 T (1000 G).
The resulting Stokes $I$ and $V$ spectra are given in Figure
\ref{fig:Gbandspectrum}, where the thick solid curves represent the
calculated $I$ and $V$ spectra taking account of both the atomic and
CH molecular line contributions, and the dash-dotted curves
represent the calculated spectra with the contribution from atomic lines
omitted.
Also plotted in this figure are the transmission curve of
a typical G-band filter (dotted curve) with a FWHM of 1 nm
and the disk centre solar intensity spectrum
        \citep[thin solid curve,][]{Kurucz1991b}.
Overall, there is good agreement between the calculated
intensity spectrum that includes all lines (thick solid curve)
and the observed spatially averaged atlas spectrum (thin solid line),
indicating that the linelists we use are adequate for our purposes.

Comparison between the solid curve and dash-dotted curves in $I$ and $V$
allows us to locate regions where the spectra are dominated by CH lines
alone.
We identified the locations at 429.74, 429.79, 430.35 through 430.41,
430.89, 431.19, and 431.35 nm as the possible regions of interest for
observations if we want to measure bright point magnetic fields
through the Zeeman effect in the CH lines.
Typical circular polarization signals at the wavelengths where
CH lines alone dominate the spectrum are 2--5\% for the 0.1 T constant
vertical field compared to 20\% for the strongest atomic lines in this
region.
These predicted values are well within the range of what could be
observed with sufficient spatial resolution.
None of these regions is dominated by a single CH line, rather
the spectrum in each is the result of overlapping lines of varying
degrees of magnetic sensitivity.
As a result, none of the profiles in Stokes $V$ has a regular
double-lobed antisymmetric shape.
In particular, the profile around 430.40 nm is notable.
The one-lobed profile is the result of two $R_{11}$ main branch lines
in the $v = 0$ band at 430.3925 and 430.3932 nm with $J_l = 1.5$ and
$J_u = 2.5$, and $g_{\mathrm{eff}} = 0.8833$ that overlap with lines that
are much less magnetically sensitive.
It raises the interesting possibility of recovering the bright point 
magnetic field with narrowband imaging in only one polarization,
avoiding the difficulties of reconstructing the polarization
map from two different exposures.
At the low rotational quantum number of the two $R_{11}$ lines the
electron spin is not completely decoupled from the inter-nuclear axis
and calculation of the effective Land\'{e} factors for these transitions
has to be treated with the intermediate case (a-b) formalism
       \citep{Schadee1978}.
Future polarimetry employing these lines would benefit from applying
a more precise value of $g_{\mathrm{eff}}$, which according to
the graph in Fig.\ 9 of 
       \citet{Berdyugina+Solanki2002}
is about 0.1 higher than the value for pure case (b).
In fact, we have verified that performing the synthesis using the
Zeeman patterns calculated in the intermediate case (a-b) by means of
the formalism developed by
       \citet{Schadee1978}
gives similar results.

\subsection{Magneto-convection snapshot}
For a prediction of the amount of Zeeman induced polarization under
more realistic conditions in the solar photosphere we solved the
Stokes transfer equations in a two-dimensional cross section
through a simulation of magneto-convection
       \citep{Stein+Bercik+Nordlund2003}.
In this simulation the physical domain extends 12 Mm in both
horizontal directions with a resolution of 96~km,
extends 2.5~Mm below the surface, and spans 3~Mm vertically.
Periodic boundary conditions were used in the horizontal directions,
and transmitting vertical boundary conditions were specified.
Radiative contributions to the energy balance were calculated by
solving radiative transfer in LTE using a four bin opacity distribution 
function.
\begin{figure}[ptbh]
  \epsscale{0.80}
  \plotone{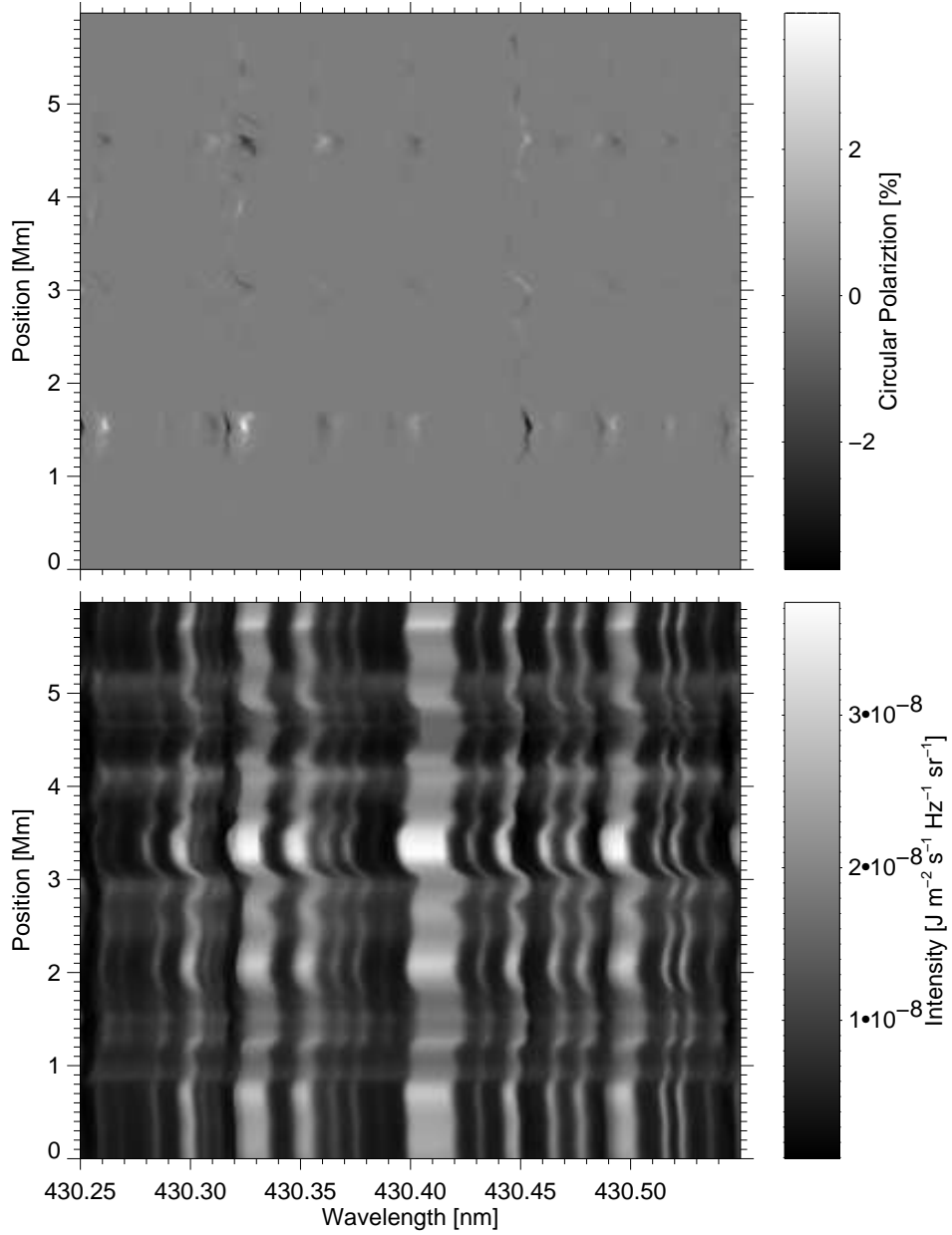}
  \caption[]{Calculated Stokes $V$ (top) and $I$ spectra for
    cross section through a magneto-convection snapshot.
    The polarization signal at wavelengths 430.36 and 430.40 is due
    the Zeeman effect in CH lines alone. 
    \label{fig:Stokes_2D}}
  \epsscale{1.0}
\end{figure}
For the Stokes transfer calculations a two-dimensional cut through
a single three-dimensional simulation snapshot was selected.
The upper part of this slice, starting at 300~km below the
photosphere, was interpolated onto a finer vertical grid, logarithmically
in densities, and linearly in temperature and velocities.
Through this interpolated cross section the Stokes
transfer equations were solved in two-dimensional geometry over
a subset of wavelengths centered on 430.4 nm.
The snapshot employed for this paper had an average vertical field
of 0.003 T.

The emergent Stokes $I$ and $V$ spectra from the slice in the
wavelength region around 430.40 nm are shown in Figure \ref{fig:Stokes_2D}.
The polarization signal at wavelengths 430.36 and 430.40 is due to
the Zeeman effect in CH lines alone.
It is most prominent at positions 1.5, 3.0, and 4.5 Mm.
All three of these locations are associated with downflows as
deduced from the redshifts in the intensity spectrum,
and are probably positioned in intergranular lanes.
At positions 1.5 and 4.5 Mm along the slice the line-of-sight
component (in the vertical direction) of the field is strongest and
results in a circular polarization signal of about 1\% of the
local continuum intensity.
Given the size of these magnetic elements of order 100 km,
and if the element is isolated without oposite magnetic
polarities within the resolution element,
they should produce a polarization signal of order at least $10^{-4}$
even when the (seeing induced) telescope resolution is only of order
1 Mm (corresponding to slightly more than $1''$).
This is within the capabilities of current polarimetric instrumentation.

It is worth noting that the spectral feature at 430.40 does not
consistently show a one-lobed profile as it does in the spectrum
produced by the one-dimensional hydrostatic model.
In particular, at position 1.5 Mm the blue lobe is visible as
well, altough much weaker than its red counterpart.
The difference with the hydrostatic model most likely derives through
an interplay between the magnetic field
and velocity gradients in the dynamic case,
and the different formation heights of the
magnetically sensitive and non-sensitive lines.
This inconsistency would make magnetic field mapping through
narrow band imaging in one polarity a less reliable method,
but its downside would have to be weighed against the improved
spatial resolution this technique promises, including the full
benefit it would have from image improving techniques like 
adaptive optics and post facto image reconstruction.

\section{Conclusion\label{sec:conclusions}}
We have solved the full set of Stokes transfer equations in two
different models of the solar atmosphere in the wavelength region
of the G band around 430 nm.
We include polarization due to the Zeeman effect in atomic transitions,
and most importantly in the molecular CH lines that are the major
contributors to opacity in the G band.
Judging from the calculated spectrum through a one-dimensional
plane parallel model of the solar atmosphere (FALC) we find several
wavelengths regions at which both the intensity and the polarization
signals are dominated by CH lines, without significant contribution
from atomic lines.
The amplitude of the calculated polarization signal in these
locations, of order 2--5\% in the hydrostatic FALC model with 0.1 T
constant vertical field and of the order of 1\% of the continuum
intensity in Stokes $V$ through the magneto-convection slice,
leads us to predict that measurement of field strength
in the CH lines that provide small-scale magnetic elements with
their characteristic brightness is a realistic prospect.

Because of the property of the lines of the strong CH main branch
to become less magnetically sensitive with increasing rotational
quantum number (as evidenced by the behavior of their effective
Land\'{e} factors as shown by
         \citet{Berdyugina+Solanki2002}) the polarization spectrum
of the G band is significantly less crowded than its intensity spectrum.
Nevertheless, overlap of lines still occurs and none of the isolated
CH features shows a regular symmetric $V$-profile even in the
absence of any flows.
In this paper we show that the effective Land\'{e} factors of the $P$ and $R$
satellite branches go asymptotically $\pm 1$ with increasing molecular
rotation (Figure \ref{fig:satellite}).
However, these satellite branch transitions are much weaker
than their counter parts in the main branches and do not
contribute significantly to the polarization signal in the G band.

Within the 3 nm wavelength interval we investigated we found one
location where several CH lines overlap and dominate the spectrum
to produce a single-lobed $V$-profile at 430.40 nm.
In particular, the spectrum calculated through the hydrostatic
FALC model shows this behavior, which prompts us to suggest that
measurement of bright-point line-of-sight magnetic field strength
could be achieved through narrowband (of order 20 pm FWHM or 200 m\AA)
imaging in one direction of circular polarization.
Such a method would have the full benifit of image quality improving
techniques like adaptive optics and phase diversity restoration
for optimal spatial resolution.
Unfortunately, the $V$ spectrum calculated through the two-dimensional
magneto-convection slice shows a very asymmetric two-lobed profile at
430.40 nm in some locations rather than a uniformly single-lobed one.
Nonetheless, the benefit of optimal spatial resolution may outweigh
the polarimetric accuracy in some cases.
Proof that such curious single-lobed $V$-profiles indeed exist
in the solar spectrum comes from recent observations of sunspots
obtained with the ZIMPOL polarimeter attached to the Gregory Coud\'e
telescope of IRSOL (locarno), which will be discussed in an upcoming paper.

We conclude that the technique of magnetic field measurement through
Stokes polarimetry of the CH lines finally holds the promise for the
development of a method that will allow us to unambiguously determine what
the role of the field is in the appearance of G-band bright points.
If such measurements are successful we will no longer have to deal
with the difficulty of interpreting magnetic field measurements
in atomic lines together with G-band intensity measurements introduced
by the different formation properties of these species.

\acknowledgements
We would like to thank the referee, Svetlana Berdyugina, for her
careful reading of the manuscript.
We are grateful to Bob Stein for providing the snapshot from which
we extracted the two-dimensional magneto-convection slice used here.
The work of AAR and JTB has been partially funded by
the Spanish Ministry of Science and Technology through project
AYA2001-1649.
This research has made use of NASA's Astrophysics Data System (ADS).



\end{document}